\documentstyle[12pt]{article}
\begin{document}

\hfill SOGANG-HEP 209/96

\hfill December 1996(revised)

\vspace{1cm}

\begin{center}
    {\large \bf  The Quantization of the Chiral Schwinger Model
    \\Based on the BFT--BFV Formalism}
\end{center}

\vspace{0.5cm}

\begin{center}
    Won T. Kim\footnote{Electronic address: wtkim@ccs.sogang.ac.kr},
    Yong-Wan Kim,
    Mu-In Park,
    and Young-Jai Park\footnote{Electronic address:
    yjpark@ccs.sogang.ac.kr} \\
    {\it Department of Physics and Basic Science Research Institute \\
     Sogang University, C.P.O. Box 1142, Seoul 100-611, Korea} \\
    {\it and} \\
    Sean J. Yoon\footnote{Electronic address: yoonsj@wm.lge.co.kr} \\
    {\it LG Electronics Research Center, Seoul 137-140, Korea}

\end{center}

\vspace{1cm}

\begin{center}
        {\bf ABSTRACT}
\end{center}

 We apply newly improved Batalin-Fradkin-Tyutin Hamiltonian method 
to the chiral Schwinger Model in the case of  
the regularization ambiguity $a>1$.
We show that one can systematically construct the first class
constraints by the BFT Hamiltonian method, and also show 
that the well-known Dirac brackets of the 
original phase space variables
are exactly the Poisson brackets of the corresponding modified fields
in the extended phase space. 
Furthermore, we show that the first class Hamiltonian
is simply obtained  by replacing the original fields 
in the canonical Hamiltonian
by these modified fields.
Performing the momentum integrations, we obtain
the corresponding first class Lagrangian in the configuration space.

\vspace{1cm}


\vspace{1cm}
\newpage

\begin{center}
\large{\bf 1. Introduction}
\end{center}

Batalin and Fradkin [1] have proposed a new kind of
quantization procedure for second class constraint systems.
When combined with the formalism of Batalin et al. (BFV) [2]
for first class constraint systems,
the BFV formalism is particularly powerful for deriving
a covariantly gauge-fixed action in configuration space.
This BFV formalism has been applied to several interesting models [3]
including the bosonized Chiral Schwinger Model (CSM) [4].
On the other hand,
Batalin-Fradkin-Tyutin (BFT) [5,6] have tried to unify the methods of
quantization for systems with only second class or both
classes of constraints, and at the same time have defined extended
algebras in terms of Poisson brackets (PB) in place of the original
ones constructed with the aid of Dirac brackets (DB) [7].

However, Banerjee, Rothe, and Rothe [8] have recently pointed out
that the work of Refs. [3,4] does not represent a systematic
application of the BFV formalism.
Banerjee et al. [9] have systematically applied the BFT
Hamiltonian method [5,6] to several second class abelian systems [10]
being led in this way to the WZW Lagrangian of the corresponding first
class theories, 
finding the new type of an abelian Wess-Zumino (WZ) action.
Very recently, we have also quantized several interesting models [11]
by using this BFT formalism. This work [8--11] is 
mainly based on the systematic, but somewhat cumbersome construction
of the first class Hamiltonian as a solution of the strongly involutive
relations with the first class constraints. In this paper,
we shall show that the desired first class Hamiltonian can also be 
obtained
from the canonical Hamiltonian by simply replacing the original fields
by modified physical fields. These modified physical fields are
obtained from the requirement of being in strong involution relation
with the first class constraints.

There has been a great progress
in the understanding of the physical meaning of
anomalies in quantum field theory through the study of the CSM.
Jackiw and Rajaraman [12]
showed that a consistent and unitary,
quantum field theory is even possible in the gauge non-invariant formulation.
Alternatively, a gauge invariant version [13] can be obtained by adding
a Wess-Zumino action to the gauge non-invariant original theory,
as was proposed by Faddeev and Shatashvili [14].
Since their works, the CSM has been still
analyzed by many authors as an archetype of anomalous gauge theory [4,8,15].
 
We shall now apply the newly improved BFT method referred to above
to the bosonized CSM.
Through the BFT analysis,
we will obtain the well-known WZ term cancelling the usual gauge anomaly
after converting the original second class system 
into the fully first class one.
The organization of the paper is as follows:
In section 2, we convert the two second class constraints of the bosonized
CSM with $a>1$ into first class constraints 
according to the BFT formalism.
In section 3, we then directly obtain the desired first class Hamiltonian
from the canonical Hamiltonian by simply replacing the original fields 
by the modified fields
and recover the DB in the extended phase space. 
In this respect, our method differs from the conventional methods [8--11].
In section 4, we find the corresponding first class Lagrangian after 
performing the momentum integrations, and
section 5 is devoted to a conclusion.

\begin{center}
\large{\bf 2. Conversion from Second to First Class Constraints}
\end{center}

In this section, we consider the bosonized CSM model in the case of $a>1$
[12,15]
\begin{equation}
    S_{CSM} ~=~ \int d^2x~\left[
                          -\frac{1}{4}F_{\mu \nu}F^{\mu \nu}
                          +\frac{1}{2}\partial_{\mu}\phi\partial^{\mu}\phi
                          +eA_{\nu}(\eta^{\mu \nu}
                          -\epsilon^{\mu \nu})\partial_{\mu}\phi
              +\frac{1}{2}ae^{2}A_{\mu}A^{\mu}~\right],
\end{equation}
where $\eta^{\mu\nu}=\mbox{diag(1,-1)}$, $\epsilon^{01}=1$, and
$a$ is a regularization ambiguity [14], which is defined for calculating
the fermionic determinant of the fermionic CSM.
The canonical momenta are given by
\begin{eqnarray}
    \pi^0~&=&~0, \nonumber \\
    \pi^1~&=&~F_{01}~=~\dot{A}_1~-~\partial_{1}A_0, \nonumber \\
    \pi_\phi~&=&~\dot{\phi}~+~e(A_0~-A_1),
\end{eqnarray}
where the overdot denotes the time derivative.
Following Dirac's standard procedure [7],
one finds one primary constraint
\begin{equation}
    \Omega_1 \equiv \pi^0 \approx 0,
\end{equation}
and one secondary constraint
\begin{equation}
    \Omega_2 \equiv \partial_1 \pi^1 + e\pi_\phi + e\partial_1 \phi
                       + e^2 A_1 + (a-1)e^2 A_0 \approx 0.
\end{equation}
This constraint is obtained by requiring the time independence of 
the primary constraint $\Omega_1$ with respect to the total Hamiltonian
\begin{equation}
    H_T = H_c + \int dx~u\Omega_1 ,
\end{equation}
where $H_c$ is the canonical Hamiltonian
\begin{eqnarray}
    H_c &=& \int\!dx~\left[~
            \frac{1}{2}(\pi^{1})^2 + \frac{1}{2}(\pi_{\phi})^2
             + \frac{1}{2}(\partial_{1}\phi)^2 - e(\pi_{\phi}
             + \partial_{1}\phi)(A_0 - A_1) \right. \nonumber \\
        &&~~~~~~ \left. - A_{0}\partial_{1}\pi^1
               - \frac{1}{2}ae^{2} \{ (A_0)^2 - (A_1)^2 \}
               + \frac{1}{2}e^{2}(A_0 - A_1)^2~\right],
\end{eqnarray}
and where denotes a Lagrange multiplier $u$.
This parameter is fixed by the second class character of the constraints
$\Omega_{i} (i=1,2)$,
which is evident from the Poisson bracktets.
Then, the constraints 
fully form the second class algebra as follows
\begin{eqnarray}
        \Delta_{ij}(x,y)
                        &\equiv&
                \{ \Omega_{i}(x), \Omega_{j}(y) \} \nonumber\\
            &=& - e^2 (a-1)\epsilon_{ij}\delta(x^{1}-y^{1}), \\
            \mbox{det} \Delta_{ij}(x,y)& \neq & 0, \nonumber
\end{eqnarray}
where $\epsilon_{12}=-\epsilon_{21}=1$.

We now introduce new auxiliary fields $\Phi^{i}$
in order to convert the second class constraint $\Omega_{i}$ into
first class ones in the extended phase space. 
Following BFT [5,6], we require these fields to satisfy
\begin{eqnarray}
   \{A^{\mu}(\mbox{or}~ \pi_{\mu}), \Phi^{i} \} &=& 0,~~~
   \{\phi(\mbox{or}~ \pi_{\phi}), \Phi^{i} \} = 0, \\ \nonumber
   \{ \Phi^i(x), \Phi^j(y) \} &=& \omega^{ij}(x,y) =
                      -\omega^{ji}(y,x),
\end{eqnarray}
where $\omega^{ij}$ is a constant and antisymmetric matrix.
Then, the strongly involutive modified constraints
$\widetilde{\Omega}_{i}$ satisfying
\begin{eqnarray}
\{\widetilde{\Omega}_{i}, \widetilde{\Omega}_{j} \}=0
\end{eqnarray}
as well as the boundary conditions,
$\widetilde{\Omega}_i \mid_{\Phi^i = 0} = \Omega_i$
are given by
\begin{equation}
  \widetilde{\Omega}_i( A^\mu, \pi_\mu, \phi, \pi_{\phi}; \Phi^j)
         =  \Omega_i + \sum_{n=1}^{\infty} \widetilde{\Omega}_i^{(n)}=0,
                       ~~~~~~\Omega_i^{(n)} \sim (\Phi^j)^n.
\end{equation}
Note that the modified constraints $\widetilde{\Omega}_{i}$
can be set strongly involutive.
The first order correction terms in the infinite series (10) are given by
[3,4]
\begin{equation}
  \widetilde{\Omega}_i^{(1)}(x) = \int dy X_{ij}(x,y)\Phi^j(y).
\end{equation}
The first class  
constraint algebra of $\widetilde{\Omega}_i$ then imposes
the following condition:
\begin{equation}
   \triangle_{ij}(x,y) +
   \int dw~ dz~
        X_{ik}(x,w) \omega^{kl}(w,z) X_{jl}(y,z)
         = 0.
\end{equation}
However, as was emphasized in Ref. [9,11], there is a natural arbitrariness
in choosing $\omega^{ij}$ and $X_{ij}$ from Eqs. (8) and (11),
which corresponds to the canonical transformation
in the extended phase space [5,6]. 

Now, making use of the above arbitrariness, 
we can take without any loss of generality the simple solutions, 
which are compatible with Eqs. (8) and (12) as
\begin{eqnarray}
  \omega^{ij}(x,y)
         &=&  \epsilon^{ij} \delta ( x^{1}-y^{1}), \nonumber  \\
  X_{ij}(x,y)
         &=&  e \sqrt{a-1} \delta_{ij} \delta ( x^{1}- y^{1}),
\end{eqnarray}
where $\omega^{ij}(x,y)$ is antisymmetric
and $X_{ij}(x,y)$ symmetric. 
Note that $X_{ij}(x,y)$ need not be in general symmetric, 
although
$\omega^{ij}(x,y)$ is always antisymmetric by the definition of Eq. (8).

With this proper choice,
the modified constraints up to first order,
\begin{eqnarray}
\widetilde{ \Omega}_{i}&=&\Omega_{i}+\Omega^{(1)}_{i} \\ \nonumber
&=&\Omega_{i} +e \sqrt{a-1} \Phi^{i},
\end{eqnarray}
form a strongly first class  constraint algebra
\begin{equation}
  \{\Omega_{i}+ \widetilde{\Omega}^{(1)}_{i},
            \Omega_{j}+ \widetilde{\Omega}^{(1)}_{j} \} = 0.
\end{equation}
The higher order correction terms in Eq. (10)
\begin{eqnarray}
\widetilde{\Omega}^{(n+1)}_{i} =- \frac{1}{n+2} \Phi^{l} 
                  \omega_{lk} X^{kj} B_{ji}^{(n)}~~~~~~~(n \geq 1)
\end{eqnarray}
with
\begin{eqnarray}
B^{(n)}_{ji} \equiv \sum^{n}_{m=0} 
       \{ \widetilde{\Omega}^{(n-m)}_{j}, 
          \widetilde{\Omega}^{(m)}_{i} \}_{(A, \pi, \phi, \pi_{\phi} )}+
         \sum^{n-2}_{m=0} 
          \{ \widetilde{\Omega}^{(n-m)}_{j}, 
              \widetilde{\Omega}^{(m+2)}_{i} \}_{(\Phi)}
\end{eqnarray}
automatically vanish as a consequence of the choice (13). 
Here, $\omega_{lk}$ and $X^{kj}$ denote
the inverse of $\omega^{lk}$ and $X_{kj}$,
and the Poisson brackets are computed using the standard canonical 
definition for $A_\mu$ and $\phi$, as well as Eq. (8).

Therefore, we have all the first class constraints
in the extended phase space
with only $\Omega_\alpha^{(1)}$ contributing in the series (10) 
as a consequence of the proper choice (13).

\begin{center}
{\large \bf 3. First Class Hamiltonian and Dirac Brackets}
\end{center}

In this section, we will show that one can easily
find the first class Hamiltonian by only
replacing the original fields in the canonical Hamiltonian with the modified
physical fields. 

Let $F \equiv (A^{\mu},\pi_{\mu},\phi, \pi_{\phi} )$ be
the variables of the original phase space. 
Then, in order to be strongly involutive in the extended phase space,
the modified physical variables
$\widetilde{F} \equiv (\widetilde{A}^{\mu},\widetilde{\pi}_{\mu},
\widetilde{\phi}, \widetilde{\pi}_{\phi} )$ are naturally required 
to be in strong involution with the first class constraints
\begin{eqnarray}
\{ \widetilde{\Omega}_{i}, \widetilde{F} \} =0
\end{eqnarray}
with
\begin{eqnarray}
\widetilde{F}(A^{\nu}, \pi_{\nu}, \phi, \pi_{\phi}; \Phi^{j} ) &=&
            F + \sum^{\infty}_{n=1} \widetilde{F}^{ (n)},
            ~~~~~~~ \widetilde{F}^{(n)} \sim (\Phi^{j})^{n}
\end{eqnarray}
satisfying the boundary conditions,
$\widetilde{F} \mid_{\Phi^i = 0} = F$.

The first order correction terms in Eq. (19) are given by
\begin{equation}
\widetilde{F}^{(1)} =  - \Phi^{j} \omega_{jk} X^{kl}
 \{ \Omega_{l}, F \}_{(A, \pi, \phi, \pi_{\phi})},
\end{equation}
or equivalently
\begin{eqnarray}
\widetilde{A}^{\mu(1)}&=&  
    ( \frac{1}{ e \sqrt{a-1}} \Phi^{2},\frac{1}{ e \sqrt{a-1}} \partial^{1} \Phi^{1} ),               \nonumber  \\
\widetilde{\pi}_{\mu}^{(1)} &=& 
     ( e \sqrt{a-1} \Phi^{1}, -\frac{e}{\sqrt{a-1}} \Phi^{1} ),\nonumber \\
\widetilde{\phi}^{(1)} &=& 
                       - \frac{1}{ \sqrt{a-1}} \Phi^{1} ,\nonumber \\
\widetilde{\pi}_{\phi}^{(1)} &=&  \frac{1}{ \sqrt{a-1}} \partial^1 \Phi^{1} .
\end{eqnarray}
Furthermore, since the above modified variables, 
up to the first order corrections,
are found to be strongly involutive
as a consequence of the proper choice (13),
the higher order correction terms
\begin{eqnarray}
\widetilde{F}^{(n+1)} &=&
                              -\frac{1}{n+1}
                              \Phi^{j}\omega_{jk} X^{kl} G^{(n)}_{l},
\end{eqnarray}
with
\begin{eqnarray}
G^{(n)}_{l} &=& \sum^{n}_{m=0}
                \{ \Omega_{i}^{(n-m)}, 
                     \widetilde{F}^{(m)}\}_{(A, \pi, \phi, \pi_{\phi} )}
                +   \sum^{n-2}_{m=0}
                \{ \Omega_{i}^{(n-m)}, \widetilde{F}^{(m+2)}\}_{(\Phi)}
           +  \{ \Omega_{i}^{(n+1)}, \widetilde{F}^{(1)} \}_{(\Phi)} 
                  \nonumber \\
\end{eqnarray}
again vanish. 
Hence, the modified variables are finally given by
\begin{eqnarray}
\widetilde{A}^{\mu} &=& A^{\mu} + \widetilde{A}^{\mu (1)} 
                     = (A^{0} + \frac{1}{e \sqrt{a-1}} \Phi^{2},
      A^{1} + \frac{1}{e \sqrt{a-1}} \partial^{1} \Phi^{1} ),  \nonumber \\
\widetilde{\pi}^{\mu} &=& \pi^{\mu} + \widetilde{\pi}^{\mu (1)} 
                      = (\pi^{0}+e \sqrt{a-1} \Phi^{1}, 
                         \pi^{1}+\frac{e}{\sqrt{a-1}} \Phi^{1} ), \nonumber \\
\widetilde{\phi} &=& \phi + \widetilde{\phi}^{(1)}  
                  = \phi - \frac{1}{ \sqrt{a-1}} \Phi^{1}, \nonumber \\
\widetilde{\pi}_\phi &=& \pi_{\phi} + \widetilde{\pi}_\phi^{(1)} 
                    = \pi_{\phi} + \frac{1}{ \sqrt{a-1}} \partial^1 \Phi^{1}.
\end{eqnarray}
Other physical quantities
depending on  $A^{\mu},~\pi_{\mu},~\phi$, and
$\pi_{\phi}$, can be also found
in principle by considering the solution as in Eqs. (18) and (19). 
However, it is expected   
that this procedure of finding these physical variables 
will in general not be simple. 

Instead of seeking such functions by solving the coupled Eqs. (22) and (23)
in the extended phase space, 
we consider a new approach using the property [6,18,19]
\begin{eqnarray}
\widetilde{K}(A^{\mu}, \pi_{\mu}, \phi, \pi_{\phi}; \Phi^{i} ) =
K(\widetilde{A}^{\mu},\widetilde{\pi}_{\mu}, \widetilde{\phi}, 
                      \widetilde{\pi}_\phi)
\end{eqnarray}
for the arbitrary function $K$ of $A^{\mu},~\pi_{\mu},~\phi$, 
and $\pi_{\phi}$.
Indeed,
\begin{eqnarray}
\{ K(\widetilde{A}^{\mu}, \widetilde{\pi}_{\mu}
, \widetilde{\phi}, \widetilde{\pi}_\phi ),\widetilde{\Omega}_{i} \}=0
\end{eqnarray}
is naturally satisfied for any function $K$ unless $K$ has 
the time derivatives.
This follows from the fact that $\widetilde{A}^{\mu}$, 
$\widetilde{\pi}_{\mu}$, $\widetilde{\phi}$, and
$\widetilde{\pi}_\phi$ already commute with
$\widetilde{\Omega}_{i}$. 
Furthermore, since the solution $K$ of Eq. (26) is
unique up to the power of the first class constraints $\widetilde{\Omega}_{i}$
[10,11,18], $K(\widetilde{A}^{\mu}, \widetilde{\pi}_{\mu}
, \widetilde{\phi}, \widetilde{\pi}_{\phi})$ can be identified
with $\widetilde{K}(A^{\mu}, \pi_{\mu}, \phi, \pi_{\phi}; \Phi)$ 
up to terms proportional to
the first class constraints $\widetilde{\Omega}_{i}$.

Now, using this elegant property,
we directly find the desired first class Hamiltonian $\widetilde{H}$
from the canonical Hamiltonian $H_{c}$ 
by only replacing the original fields with the modified fields:
\begin{eqnarray}
   \widetilde{H}(A^{\mu}, \pi_{\nu},\phi, \pi_{\phi}; \Phi^{i} )
      \!\!&=&\!\! H_{c}(\widetilde{A}^{\mu}, \widetilde{\pi}_{\nu}
      , \widetilde{\phi},\widetilde{ \pi}_{\phi}) \nonumber \\
  \!\!&=&\!\! H_{c}(A^{\mu}, \pi_{\nu},\phi, \pi_{\phi})  
       +\int\! dx
      \left[
         \frac{1}{2}(\partial_1 \Phi^1)^2
         + \frac{e^2}{2(a-1)}(\Phi^1)^2 
         + \frac{1}{2}(\Phi^2)^2  \right. \nonumber \\
&& \left. 
      + \frac{1}{e\sqrt{a-1}}  [e^2 \pi^1 - e^2(a-1) \partial_1 A_1 ]
          \Phi^1 - \frac{1}{e\sqrt{a-1}}\widetilde{\Omega}_2 \Phi^2
      \right].
\end{eqnarray}
By construction, $\widetilde{H}$ is strongly involutive 
with the first class constraints (14),
\begin{eqnarray}
\{ \widetilde{\Omega}_{i}, \widetilde{H} \} =0.
\end{eqnarray}

In this way, all second class constraints 
$\Omega_{i}(A_{\mu}, \pi^{\mu}, \phi, \pi_{\phi} ) \approx 0$ 
are possible to convert into
the first class ones $\widetilde{\Omega}_{i}(A_{\mu},
\pi^{\mu}, \phi, \pi_{\phi}; \Phi ) = 0$ 
satisfying the boundary conditions
$\widetilde{\Omega}_{i} |_{\Phi=0}=\Omega_{i}$.
It is interesting to note that the difference
$\widetilde{H}(A_\mu,\Pi^\mu,\phi, \pi_{\phi};  \Phi^{i})-
    H_c(A_\mu, \pi^\mu,\phi, \pi_{\phi}) $ is  independent of the bosonized
    matter field $\phi$ and the momentum $\pi_{\phi}$.
The modified constraints trivially satisfy  
this property,
$\widetilde{\Omega}_{i}(A^{\mu}, \pi_{\mu}, \phi, \pi_{\phi}; \Phi)=
\Omega_{i}(\widetilde{A}^{\mu},\widetilde{\pi}_{\mu}, \widetilde{\phi},
\widetilde{\pi}_\phi)$. 

Since in the Hamiltonian formalism the first class constraint system
indicates the presence of a local symmetry,
this  completes  the  operatorial conversion of the original
second class system with Hamiltonian $H_c$ and constraints $\Omega_i$
into the first class one with Hamiltonian $\widetilde H$ and constraints
$\widetilde{\Omega}_i$ by using the property (25).

On the other hand, in Dirac
formulation [7] one can directly deal with 
the second class constraints system
by using the DB, rather than extending the phase space.
Thus, it seems that the Dirac and BFT formalisms 
are drastically different ones. 
However, one can recognize that the DB can be automatically read off 
from the BFT formalism [5,6] by noting
\begin{eqnarray}
\{A, B \}_{D} &\equiv &\{ \widetilde{A}, \widetilde{B} \} |_{\Phi=0}  
                                 \nonumber \\
              &=&\{A, B \}-\{A, \Omega_{k} \} \Delta^{kk'} \{\Omega_{k'}, B \},
\end{eqnarray}
where $\Delta^{kk'}=-X^{lk} \omega_{ll'} X^{l'k'}$ is the inverse of
$\Delta_{kk'}$ in Eq. (7), since this bracket has the property
\begin{eqnarray}
\{ \Omega_{i}, A \}_{D} =0, ~~
\{ \Omega_{i}, \Omega_{j} \}_{D} = 0.
\end{eqnarray}

Now, let us explicitly consider the brackets 
between the phase space variables in Eq. (24) for the CSM.
The non-vanishing results are as follows
\begin{eqnarray}
&& \{\widetilde{A}_{1}(x), \widetilde{\pi}^{1}(y) \} |_{\Phi=0}
    = \{\widetilde{A}_{1}(x), \widetilde{\pi}^{1}(y) \} 
    = \delta(x^{1}-y^{1}),  
      \nonumber \\
&&  \{\widetilde{\phi}(x), \widetilde{\pi}_\phi (y)\} |_{\Phi=0}
    = \{\widetilde{\phi}(x), \widetilde{\pi}_\phi (y) \} 
    = \delta(x^{1}-y^{1}),   
      \nonumber  \\
&& \{\widetilde{A}_0 (x), \widetilde{A}_{1}(y) \} |_{\Phi=0}
    = \{\widetilde{A}_0 (x), \widetilde{A}_{1}(y) \} 
    = - \frac{1}{e^2 (a-1) } \partial^1_x \delta(x^1-y^1),  
      \nonumber \\
&& \{\widetilde{A}_0 (x), \widetilde{\pi}^1 (y) \} |_{\Phi=0}
    = \{\widetilde{A}_0 (x), \widetilde{\pi}^1 (y) \} 
    = -\frac{1}{(a-1)} \delta (x^1-y^1),  
     \nonumber \\
&&\{\widetilde{A}_0 (x), \widetilde{\phi}(y) \} |_{\Phi=0}
    = \{\widetilde{A}_0 (x), \widetilde{\phi}(y) \} 
    = \frac{1}{e (a-1)} \delta(x^1-y^1 ),  
     \nonumber \\
&&\{\widetilde{A}_0 (x), \widetilde{\pi}_\phi (y) \} |_{\Phi=0}
    = \{\widetilde{A}_0 (x), \widetilde{\pi}_\phi (y) \} 
    = \frac{1}{e (a-1)} \partial^1_x \delta(x^1-y^1)
\end{eqnarray}
due to the linear correction 
(i.e., first order correction only for the CSM) of the
modified fields $\widetilde{A}^{\mu}$, $\widetilde{\pi}_{\mu}$,
$\widetilde{\phi}$, and $\widetilde{\pi}_\phi$. 
These brackets are exactly the same
as the results of the usual Dirac ones [15]. 
Hence, for the case of the CSM, we show that 
the DB of the fields in the original phase space
are exactly same as the Poisson brackets of 
the corresponding modified fields in the extended space.
This is in agreement with the results of Ref. [16], as it should,
since the ``tilde'' variables correspond to the gauge invariant variables
of that reference.

\begin{center}
{\large \bf 4. Corresponding First Class Lagrangian}
\end{center}

In this section, we consider the partition function of the model
in order to present the Lagrangian corresponding to $\widetilde{H}$
in the canonical Hamiltonian formalism.
As a result, we will unravel the correspondence of
the Hamiltonian approach with
the well-known St\"uckelberg's formalism.

First, let us identify the new variables
$\Phi^i$ as a canonically conjugate pair
($\theta$, $\pi_\theta$) in the Hamiltonian formalism
\begin{eqnarray}
\Phi^{i} \equiv \sqrt{a-1} (\theta, \frac{1}{a-1} \pi_{\theta} )
\end{eqnarray}
satisfying Eqs. (8) and (13).
Then, the starting phase space partition function is given by the Faddeev
formula [17] as follows
\begin{equation}
Z=  \int  {\cal D} A^\mu
          {\cal D} \pi_\mu
          {\cal D} \phi
          {\cal D} \pi_{\phi}
          {\cal D} \theta
          {\cal D} \pi_\theta
               \prod_{i,j = 1}^{2} \delta(\widetilde{\Omega}_i)
                           \delta(\Gamma_j)
                \mbox{det} \mid \{ \widetilde{\Omega}_i, \Gamma_j \} \mid
                e^{iS},
\end{equation}
where
\begin{equation}
S  =  \int d^2x \left(
               \pi_\mu {\dot A}^\mu +\pi_{\phi} {\dot \phi} + \pi_\theta {\dot \theta} - \widetilde {\cal H}
                \right),
\end{equation}
with the Hamiltonian density $\widetilde {\cal H}$ corresponding to the 
Hamiltonian
$\widetilde H$ of Eq. (27), which is now expressed in terms
of $(\theta, \pi_\theta)$ instead of $\Phi^i$.
The gauge fixing conditions $\Gamma_i$ are chosen
so that the determinant occurring in
the functional measure is nonvanishing.
Moreover, $\Gamma_i$ may be assumed to be independent of the momenta
so that these are considered as Faddeev-Popov type gauge conditions
[8--11].

Before performing the momentum integrations to obtain the
partition function in the configuration space,
it seems appropriate to comment on the strongly involutive Hamiltonian (27).
If we use this Hamiltonian,
we do not naturally generate
the first class Gauss' law constraint $\widetilde{\Omega}_2$ from
the time evolution of the primary constraint $\widetilde{\Omega}_1$.
Therefore, in order to arrive at familiar results,
we use an equivalent first class Hamiltonian,
which differs from the Hamiltonian (27)
by adding a term proportional to the first class constraint
$\widetilde{\Omega}_2$ as follows
\begin{equation}
\widetilde{H}' = \widetilde H + \int d x \frac{1}{e (a-1)} \pi_{\theta} \widetilde{\Omega}_2.
\end{equation}
Then, the modified Hamiltonian $\widetilde{H}'$ naturally generates the
Gauss' law constraint such that
\begin{eqnarray}
\{ \widetilde{\Omega}_1, \widetilde{H}'\} &=& \widetilde{\Omega}_2, \nonumber \\
\{ \widetilde{\Omega}_2, \widetilde{H}'\} &=& 0.
\end{eqnarray}
Note that $\widetilde{H}$ and $\widetilde{H}'$ act in the same way
on physical states, since such states are
annihilated by the first class constraints.
Similarly, the equations of motion for observables ({\it i.e.}
gauge invariant variables) will also be unaffected by this difference
since $\widetilde{\Omega}_2$ can be regarded as
the generator of the gauge transformations.

Next, we perform the momentum integrations to obtain the
configuration space partition function.
The  $\pi^0$ integration is trivially performed by exploiting
the delta function
$~~\delta(\widetilde{\Omega}_1)~ =~ \delta[\pi^0 + e(a-1)\theta]$ .
Then, after exponentiating the remaining delta function $
\delta(\widetilde{\Omega}_2) =
    \delta[ \partial_1 \pi^1 + e\pi_\phi + e\partial_1 \phi
            + e^2 A_1 + (a-1)e^2 A_0 + e\pi_\theta ] $ 
in terms of a
Fourier variable $\xi$ as $\delta(\widetilde{\Omega}_2)=
\int{\cal D}\xi e^{-i\int d^2x\xi\tilde{\Omega}_2}$, 
transforming $A_0 \to A_0 + \xi$,
and integrating the other momentum variables $\pi_\phi$ and $\pi^1$,
we obtain the following intermediate action
\begin{eqnarray}
    S &=& S_{\rm CSM}   \nonumber \\
       &+& \int d^2x \left[
      \theta \{ (a-1)e ( \partial_1 A_1 - \dot A_0 - \dot \xi )
              + \frac{1}{2}(a-1) \partial_1^2 \theta
              -e \epsilon^{\mu\nu} \partial_\mu A_\nu \} \right.
                                       \nonumber \\
       &+&    \left.
              \pi_\theta \{ \dot\theta -e\xi-\frac{1}{2(a-1)}\pi_\theta \}
              - \frac{1}{2} (a-1)e^2 \xi^2
              \right],
\end{eqnarray}
and corresponding integration measure as given by
\begin{equation}
[{\cal D} \mu] = {\cal D} A_\mu
         {\cal D} \phi
                 {\cal D} \theta
         {\cal D} \pi_\theta
                 {\cal D} \xi
                 \prod^2_{\beta = 1}
           \delta\left(\Gamma_{\beta}[A_0 + \xi, A_1, \theta]\right)
        \det \mid \{\widetilde{\Omega}_{\alpha}, \Gamma_{\beta}\} \mid.
\end{equation}

It seems appropriate to comment on the unitary gauge fixing.
At this stage, the original theory is reproduced
in the usual unitary gauge corresponding to the choice
\begin{equation}
    \Gamma_\alpha = ( \theta, \pi_\theta),
\end{equation}
in Eq. (38).
Note that this gauge fixing is self-consistent since 
with the gauge fixing condition $\theta \approx 0$,
the condition $\pi_\theta \approx 0$ is naturally generated from
the time evolution of $\theta$, {\it i.e.},
$\dot\theta = \{\theta, \tilde{H}' \}
= \frac{1}{(a-1)} \pi_\theta \approx 0$.
One thus recognizes that
the new fields $\Phi^i$ are nothing but the gauge degrees of
freedom, which can be removed by gauge transformation.

Finally, we perform the Gaussian integration over $\pi_\theta$.
Then, all terms including $\xi$ in the action are found to cancel,
the resulting action being given by
\begin{eqnarray}
    S &=& S_{CSM} + S_{WZ} ~;           \nonumber \\
    S_{WZ} &=&  \int d^2x \left[
                \frac{1}{2}(a-1)\partial_\mu \theta \partial^\mu \theta
                - e \theta \{ (a-1)\eta^{\mu\nu} + \epsilon^{\mu\nu} \}
                    \partial_{\mu} A_\nu  \right],
\end{eqnarray}
where $S_{WZ}$ is the well-known WZ term, which serves to cancel the
gauge anomaly.
The corresponding Liouville measure now reads
\begin{equation}
[{\cal D} \mu] = {\cal D} A_\mu
         {\cal D} \phi
                 {\cal D} \theta
                 {\cal D} \xi
                 \prod^2_{\beta = 1}
           \delta\left(\Gamma_{\beta}[A_0 + \xi, A_1, \theta]\right)
        \det \mid \{\widetilde{\Omega}_{\alpha}, \Gamma_{\beta}\} \mid.
\end{equation}
Starting from the Lagrangian (40),
we easily reproduce all the first class constraints
(14) and the modified Hamiltonian (35), which is
effectively equivalent to the strongly involutive Hamiltonian (27).

Since the quantization based on the BFV formalism [2] is by now standard and
straightforward, it seems appropriate to just comment on the BRST gauge-fixed
covariant action. After introducing ghosts, antighosts, their conjugate 
momenta along with auxiliary fields, and fixing the gauge functions,
we obtain, after performing as usual the {\it momentum} integrations 
in the generating functional,
a standard BRST gauge-fixed covariant action [4,8] including the
well-known Wess-Zumino scalar in configuration space.
On the other hand, if instead we perform the integration 
over the ghost fields, rather than their momenta, 
we obtain the nonlocal action and 
nonstandard BRST transformation proposed by several
authors [20]. 
Furthermore, one easily verifies that the local and the
nonlocal transformation are simply related 
by a change of variables, so that
the nonlocal action and the corresponding nonlocal BRST transformation are
in fact equivalent to the local theory [21].

\begin{center}
\large{\bf 5. Conclusion}
\end{center}

We have quantized the bosonized CSM with $a>1$
by using the newly improved BFT--BFV formalism.
We have shown that one can systematically construct the first class
constraints, and established the relation between the DB defined 
in the original second class system and the Poisson Brackets
defined in the extended phase space.
We have
directly obtained the first class involutive Hamiltonian
from the canonical Hamiltonian by simply replacing the original fields 
by the modified fields, and 
found the corresponding first class Lagrangian
including well-known WZ terms.
It will be interesting to apply this
improved BFT-BFV method to non-abelian cases,
including the non-abelian Proca model [22], 
which seem to be very difficult to analyze within the framework of 
the original BFT--BFV formalism [9--11].

\begin{center}
\section*{Acknowledgments}
\end{center}

The present study was supported in part by
the Sogang University Research Grants in 1996, and
the Basic Science Research Institute Program,
Ministry of Education, Project No. {\bf BSRI}-96-2414.

\vspace{2cm}

\section*{References}

\begin{description}{}
\item{[1]} Batalin I A and Fradkin E S 1986 
           {\it Phys. Lett.} {\bf B180} 157
\item{[2]} Fradkin E S and Vilkovisky G A 1975 
           {\it Phys. Lett.} {\bf 55B} 224
\item{}\hspace{0.45cm} Batalin I A and Vilkovisky G A 1977 
          {\it Phys. Lett.} {\bf 69B} 309
\item{[3]} Fujiwara T, Igarashi Y and Kubo J 1990
           {\it Nucl. Phys.} {\bf B341} 695
\item{[4]} Moshe M and Oz Y 1989 
          {\it Phys. Lett.} {\bf B224} 145
\item{}\hspace{0.45cm} Kim Y W {\it et al} 1992 
          {\it Phys. Rev.} {\bf D46} 4574
\item{}\hspace{0.45cm} Zhou J G, Miao Y G, and Liu Y Y 1994 
          {\it J. Phys.} {\bf G20} 35
\item{[5]} Batalin I A and Fradkin E S 1987 
           {\it Nucl. Phys.} {\bf B279}  514
\item{[6]} Batalin I A and Tyutin I V 1991 
           {\it Int. J. Mod. Phys.} {\bf A6} 3255
\item{}\hspace{0.45cm} Fradkin E S 1988 
          {\it Lecture of the Dirac Model of ICTP}
          Miramare-Trieste 
\item{[7]} Dirac P A M 1964 {\it Lectures on quantum mechanics} 
           New York, Yeshiba University Press 
\item{[8]} Banerjee R, Rothe H J and Rothe K D 1994
           {\it Phys. Rev.} {\bf D49}  5438
\item{[9]} Banerjee R 1993
           {\it Phys. Rev.} {\bf D48} R5467
\item{[10]} Banerjee N, Ghosh S and Banerjee R 1994
           {\it Nucl. Phys.} {\bf B417} 257
\item{}\hspace{0.6cm} Banerjee R, Rothe H J and Rothe K D 1994
           {\it Nucl. Phys.} {\bf B426} 129
\item{[11]} Kim W T and Park Y J 1994 
           {\it Phys. Lett.} {\bf B336} 376
\item{}\hspace{0.6cm} Kim Y W {\it et al} 1995
           {\it Phys. Rev.} {\bf D51} 2943
\item{[12]} Jackiw R and Rajaraman R 1985
            {\it Phys. Rev. Lett.} {\bf 54} 1219;
            1985 {\it Phys. Rev. Lett.} {\bf 54} 2060(E)
\item{}\hspace{0.6cm} Falck N K and Kramer G 1987 
            {\it Ann. Phys.} {\bf 176} 330
\item{}\hspace{0.6cm} Shizuya K 1988 
           {\it Phys. Lett.} {\bf B213} 298
\item{}\hspace{0.6cm} Harada K 1990 
           {\it Phys. Rev. Lett.} {\bf 64} 139
\item{[13]} Schaposnik F and  Viallet C 1986
           {\it Phys. Lett.} {\bf B177} 385
\item{}\hspace{0.6cm} Harada K and Tsutsui I 1987 
           {\it Phys. Lett.} {\bf B183} 311
\item{[14]} Faddeev L D and Shatashvili S S 1986
           {\it Phys. Lett.} {\bf B167} 225
\item{[15]} Rajaraman R 1985 
            {\it Phys. Lett.} {\bf B154} 305
\item{}\hspace{0.6cm} Girotti H O, Rothe H J and Rothe K D 1986
            {\it Phys. Rev.} {\bf D34} 592
\item{}\hspace{0.6cm} Rothe H J and Rothe K D 1989 
            {\it Phys. Rev.} {\bf D40} 545
\item{}\hspace{0.6cm} Sladkowski J 1992 
            {\it Phys. Lett.} {\bf B296} 361
\item{}\hspace{0.6cm} Zhou J G, Miao Y G and Liu Y Y 1994
            {\it Mod. Phys. Lett.} {\bf A9} 1273
\item{[16]} Rothe K D and Girotti H O 1989 
            {\it Int. J. Mod. Phys.} {\bf A4} 3041
\item{[17]} Faddeev L D and Popov V N 1967
            Phys. Lett. B25 29
\item{[18]} Kim Y W {\it et al} 1996
            {\it Int. J. Mod. Phys.} {\bf A} (to be published)
\item{[19]} Amorim R 1995
            {\it Z. Phys.} {\bf C67} 695
\item{}\hspace{0.6cm} Banerjee N and Banerjee R 1995
            hep-th/9511212
\item{[20]}  Lavelle M and McMullan D 1993 
            {\it Phys. Rev. Lett.} {\bf 71} 3758
\item{}\hspace{0.6cm} Tang Z and Finkelstein D 1994 
            {\it Phys. Rev. Lett.} {\bf 73} 3055  
\item{}\hspace{0.6cm} Rabello S J and Gaete P 1995
           {\it Phys. Rev.} {\bf D52}, 7205
\item{}\hspace{0.6cm} Shin H {\it et al} 1996 
           {\it J. Korean Phys. Soc.} {\bf 29} 392
\item{[21]} Rivelles V O 1995 {\it Phys. Rev. Lett.} 
             {\bf 75} 4150; 1996 {\it Phys. Rev.} {\bf D53} 3247 
\item{[22]} Banerjee N, Banerjee R and Ghosh S 1995 
            {\it Ann. Phys.} {\bf 241} 237
\end{description}
\end{document}